\documentclass[twocolumn]{aastex62}
\pdfoutput=1 
\usepackage{amsmath,amstext}
\usepackage[T1]{fontenc}
\usepackage[figure,figure*]{hypcap}
\graphicspath{{./}{figures/}}

\received{?}
\revised{?}
\accepted{?}
\submitjournal{AAS}

\shorttitle{TKS III: A Stellar Obliquity Measurement of TOI-1726 c}
\shortauthors{Dai et al.}

\begin{document}

\title{The TESS-Keck Survey III: A Stellar Obliquity Measurement of TOI-1726 c}

\author[0000-0002-8958-0683]{Fei Dai}
\affiliation{Division of Geological and Planetary Sciences,
California Institute of Technology,
1200 East California Blvd,
Pasadena, CA 91125, USA}
\email{fdai@caltech.edu}

\author[0000-0001-8127-5775]{Arpita Roy}
\affiliation{Department of Astronomy, California Institute of Technology, Pasadena, CA 91125, USA}

\author[0000-0003-3504-5316]{Benjamin Fulton}
\affiliation{NASA Exoplanet Science Institute/Caltech-IPAC, MC 314-6, 1200 E California Blvd, Pasadena, CA 91125, USA}

\author[0000-0003-0149-9678]{Paul Robertson}
\affiliation{Department of Physics \& Astronomy, The University of California, Irvine, Irvine, CA 92697, USA}

\author[0000-0001-8058-7443]{Lea Hirsch}
\affiliation{Kavli Institute for Particle Astrophysics and Cosmology, Stanford University, Stanford, CA, USA}

\author[0000-0002-0531-1073]{Howard Isaacson}
\affiliation{501 Campbell Hall, University of California at Berkeley, Berkeley, CA 94720, USA}

\author[0000-0003-1762-8235]{Simon Albrecht}
\affiliation{Stellar Astrophysics Centre, Department of Physics and Astronomy, Aarhus University, Ny Munkegade 120, DK-8000 Aarhus C, Denmark}

\author[0000-0003-3654-1602]{Andrew W. Mann}%
\affiliation{Department of Physics and Astronomy, The University of North Carolina at Chapel Hill, Chapel Hill, NC 27599, USA} 

\author[0000-0002-2607-138X]{Martti H. Kristiansen}%
\affiliation{Brorfelde Observatory, Observator Gyldenkernes Vej 7, DK-4340 T\o{}ll\o{}se, Denmark}
\affiliation{DTU Space, National Space Institute, Technical University of Denmark, Elektrovej 327, DK-2800 Lyngby, Denmark}

\author[0000-0002-7030-9519]{Natalie M. Batalha}
\affiliation{Department of Astronomy and Astrophysics, University of California, Santa Cruz, CA 95060, USA}

\author[0000-0001-7708-2364]{Corey Beard}
\affiliation{Department of Physics \& Astronomy, The University of California, Irvine, Irvine, CA 92697, USA}

\author[0000-0003-0012-9093]{Aida Behmard}
\affiliation{Division of Geological and Planetary Sciences,
California Institute of Technology,
1200 East California Blvd,
Pasadena, CA 91125, USA}

\author[0000-0003-1125-2564]{Ashley Chontos}
\altaffiliation{NSF Graduate Research Fellow}
\affiliation{Institute for Astronomy, University of Hawai`i, 2680 Woodlawn Drive, Honolulu, HI 96822, USA}

\author{Ian J.\ M.\ Crossfield}
\affiliation{Department of Physics and Astronomy, University of
  Kansas, Lawrence, KS, USA}

\author[0000-0002-4297-5506]{Paul A.\ Dalba}
\altaffiliation{NSF Astronomy and Astrophysics Postdoctoral Fellow}
\affiliation{Department of Earth and Planetary Sciences, University of California, Riverside, CA 92521, USA}

\author[0000-0001-8189-0233]{Courtney Dressing}
\affiliation{501 Campbell Hall, University of California at Berkeley, Berkeley, CA 94720, USA}

\author[0000-0002-8965-3969]{Steven Giacalone}
\affiliation{501 Campbell Hall, University of California at Berkeley, Berkeley, CA 94720, USA}

\author[0000-0002-0139-4756]{Michelle Hill}
\affiliation{Department of Earth and Planetary Sciences, University of California, Riverside, CA 92521, USA}

\author[0000-0001-8638-0320]{Andrew W. Howard}
\affiliation{California Institute of Technology, Pasadena, CA 91125, USA}

\author[0000-0001-8832-4488]{Daniel Huber}
\affiliation{Institute for Astronomy, University of Hawai`i, 2680 Woodlawn Drive, Honolulu, HI 96822, USA}

\author[0000-0002-7084-0529]{Stephen R. Kane}
\affiliation{Department of Earth and Planetary Sciences, University of California, Riverside, CA 92521, USA}

\author[0000-0002-6115-4359]{Molly Kosiarek}
\affiliation{Department of Astronomy and Astrophysics, University of California, Santa Cruz, CA 95060, USA}

\author[0000-0001-8342-7736]{Jack Lubin}
\affiliation{Department of Physics \& Astronomy, The University of California, Irvine, Irvine, CA 92697, USA}

\author[0000-0002-7216-2135]{Andrew Mayo}
\affiliation{501 Campbell Hall, University of California at Berkeley, Berkeley, CA 94720, USA}

\author[0000-0003-4603-556X]{Teo Mocnik}
\affiliation{Gemini Observatory Northern Operations, 670 N. A'ohoku Place, Hilo, HI 96720, USA}

\author[0000-0001-8898-8284]{Joseph M. Akana Murphy}
\altaffiliation{NSF Graduate Research Fellow}
\affiliation{Department of Astronomy and Astrophysics, University of California, Santa Cruz, CA 95060, USA}

\author[0000-0003-0967-2893]{Erik A. Petigura}
\affil{Department of Physics \& Astronomy, University of California Los Angeles, Los Angeles, CA 90095, USA}

\author[0000-0001-8391-5182]{Lee Rosenthal}
\affiliation{Department of Astronomy, California Institute of Technology, Pasadena, CA 91125, USA}

\author[0000-0003-3856-3143]{Ryan A. Rubenzahl}
\altaffiliation{NSF Graduate Research Fellow}
\affiliation{Department of Astronomy, California Institute of Technology, Pasadena, CA 91125, USA}

\author{Nicholas Scarsdale}
\affiliation{Department of Astronomy and Astrophysics, University of California, Santa Cruz, CA 95060, USA}

\author[0000-0002-3725-3058]{Lauren M. Weiss}
\affiliation{Institute for Astronomy, University of Hawai`i, 2680 Woodlawn Drive, Honolulu, HI 96822, USA}

\author[0000-0002-4290-6826]{Judah Van Zandt}
\affil{Department of Physics \& Astronomy, University of California Los Angeles, Los Angeles, CA 90095, USA}

\author[0000-0003-2058-6662]{George R. Ricker}
\affiliation{Department of Physics and Kavli Institute for Astrophysics and Space Research, Massachusetts Institute of Technology, Cambridge, MA 02139, USA}

\author[0000-0001-6763-6562]{Roland Vanderspek}
\affiliation{Department of Physics and Kavli Institute for Astrophysics and Space Research, Massachusetts Institute of Technology, Cambridge, MA 02139, USA}

\author[0000-0001-9911-7388]{David W. Latham}
\affiliation{Center for Astrophysics | Harvard \& Smithsonian, 60 Garden St, Cambridge, MA 02138, USA}

\author[0000-0002-6892-6948]{Sara Seager}
\affiliation{Department of Physics and Kavli Institute for Astrophysics and Space Research, Massachusetts Institute of Technology, Cambridge, MA
02139, USA}
\affiliation{Department of Earth, Atmospheric and Planetary Sciences, Massachusetts Institute of Technology, Cambridge, MA 02139, USA}
\affiliation{Department of Aeronautics and Astronautics, MIT, 77 Massachusetts Avenue, Cambridge, MA 02139, USA}

\author[0000-0002-4265-047X]{Joshua N. Winn}
\affiliation{Department of Astrophysical Sciences, Princeton University, 4 Ivy Lane, Princeton, NJ 08544, USA}

\author[0000-0002-4715-9460]{Jon M. Jenkins}
\affiliation{NASA Ames Research Center, Moffett Field, CA, 94035, USA}

\author[0000-0003-1963-9616]{Douglas A. Caldwell}
\affiliation{SETI Institute, Mountain View, CA, USA}
\affiliation{NASA Ames Research Center, Moffett Field, CA, 94035, USA}

\author[0000-0002-9003-484X]{David Charbonneau}
\affiliation{Center for Astrophysics | Harvard \& Smithsonian, 60 Garden St, Cambridge, MA 02138, USA}

\author[0000-0002-6939-9211]{Tansu Daylan}
\affiliation{Department of Physics and Kavli Institute for Astrophysics and Space Research, Massachusetts Institute of Technology, 70 Vassar Street, Cambridge, MA 02139,
USA}
\affiliation{Kavli Fellow}

\author[0000-0002-3164-9086]{Maximilian N. G{\"u}nther}
\affiliation{Department of Physics and Kavli Institute for Astrophysics and Space Research, Massachusetts Institute of Technology, 70 Vassar Street, Cambridge, MA 02139,
USA}
\affiliation{Juan Carlos Torres Fellow}

\author[0000-0003-1447-6344]{Edward Morgan}
\affiliation{Department of Aeronautics and Astronautics, MIT, 77 Massachusetts Avenue, Cambridge, MA 02139, USA}

\author[0000-0002-8964-8377]{Samuel N. Quinn}
\affiliation{Center for Astrophysics | Harvard \& Smithsonian, 60 Garden St, Cambridge, MA 02138, USA}

\author[0000-0003-4724-745X]{Mark E. Rose}
\affiliation{NASA Ames Research Center, Moffett Field, CA, 94035, USA}

\author[0000-0002-6148-7903]{Jeffrey C. Smith}
\affiliation{NASA Ames Research Center, Moffett Field, CA, 94035, USA}
\affiliation{SETI Institute, Mountain View, CA, USA}



\begin{abstract}

We report the measurement of a spectroscopic transit of TOI-1726 c, one of two planets transiting a  G-type star with $V$ = 6.9 in the Ursa Major Moving Group ($\sim$400 Myr). With a precise age constraint from cluster membership, TOI-1726 provides a great opportunity to test various obliquity excitation scenarios that operate on different timescales. By modeling the Rossiter-McLaughlin (RM) effect, we derived a sky-projected obliquity of  $-1^{+35}_{-32}~^{\circ}$. This result rules out a polar/retrograde orbit; and is consistent with an aligned orbit for planet c. Considering the previously reported, similarly prograde RM measurement of planet b and the transiting nature of both planets, TOI-1726 tentatively conforms to the overall picture that compact multi-transiting planetary systems tend to have coplanar, likely aligned orbits. TOI-1726 is also a great atmospheric target for understanding differential atmospheric loss of sub-Neptune planets (planet b 2.2 $R_\oplus$ and c 2.7 $R_\oplus$ both likely underwent photoevaporation). The coplanar geometry points to a dynamically cold history of the system that simplifies any future modeling of atmospheric escape.
\end{abstract}




\section{Introduction} \label{sec:intro}
The stellar obliquity is the angle between the rotation axis of the host star and the normal of the orbital plane of its planet. While the planets in the Solar System are well-aligned with the Sun (obliquity $\lesssim7^{\circ}$), many of the known exoplanets have polar or even retrograde orbits \citep[e.g.][]{Kepler63,HD3167}. These spin-orbit misalignments are often interpreted as signposts of a dynamically hot formation or evolution history. Various mechanisms have been proposed to be responsible for tilting the orbits of planets. Many of these mechanisms operate on different timescales: primordial disk misalignment during the disk-hosting stage \citep[$\lesssim 3$ Myr, e.g.][]{Lai,Batygin}; nodal procession induced by an inclined companion \citep[$\sim 3.5$ Myr for HAT-P-11b,][]{Yee}; the Kozai-Lidov mechanism operates on a wide range of timescales 10$^4$ to $10^8$ yr depending on the system configuration \citep[e.g.][]{Fabrycky}; and secular chaos between longer-period giant planets can happen in 10$^7$ to $10^8$ yr \citep[e.g.][]{WuLithwick}. A sample of obliquity measurements spanning a range of precise host star ages will help us distinguish these orbit-tilting mechanisms.

Precise stellar ages for main sequence stars are hard to come by, particularly for later-type stars which barely evolve over a Hubble time. Our best age constraints come from establishing cluster membership of a planet host so that the ensemble study of kinematics, stellar activity, Li abundance, gyrochronology and isochronal fitting of other stars in the same cluster can firmly pin down the stellar age. So far, there are about a dozen planet hosts found in young clusters \citep[e.g.][]{David,Mann}. They are crucial for our understanding of various aspects of planet formation and evolution. TOI-1726 is a G-type star in the Ursa Major Moving Group \citep[414$\pm23$ Myr,][]{Jones} that hosts two transiting sub-Neptune planets with 2.2 and 2.7 $R_\oplus$ on 7 and 20-day orbits \citep{Mann2020}. With a $V$-band magnitude of 6.9 and a projected rotational velocity $v$sin$i$ of $\sim 7$ km/s, TOI-1726 provides a rare opportunity to measure the stellar obliquity of a young sub-Neptune planet. In this work, we discuss a new measurement of the stellar obliquity of planet c. 
 
This letter is structured as follows. In Section 2 we present the spectroscopic measurements of the TOI-1726. Section 3 describes the constraints on the stellar parameters using both spectroscopy and Gaia information. In Section 4, we present a joint analysis of the {\it TESS} light curve and the Rossiter-McLaughlin (RM) effect to measure the stellar obliquity of TOI-1726 c. Section 5 discusses the implication of our finding.

\begin{figure*}
\begin{center}
\includegraphics[width = 1.8\columnwidth]{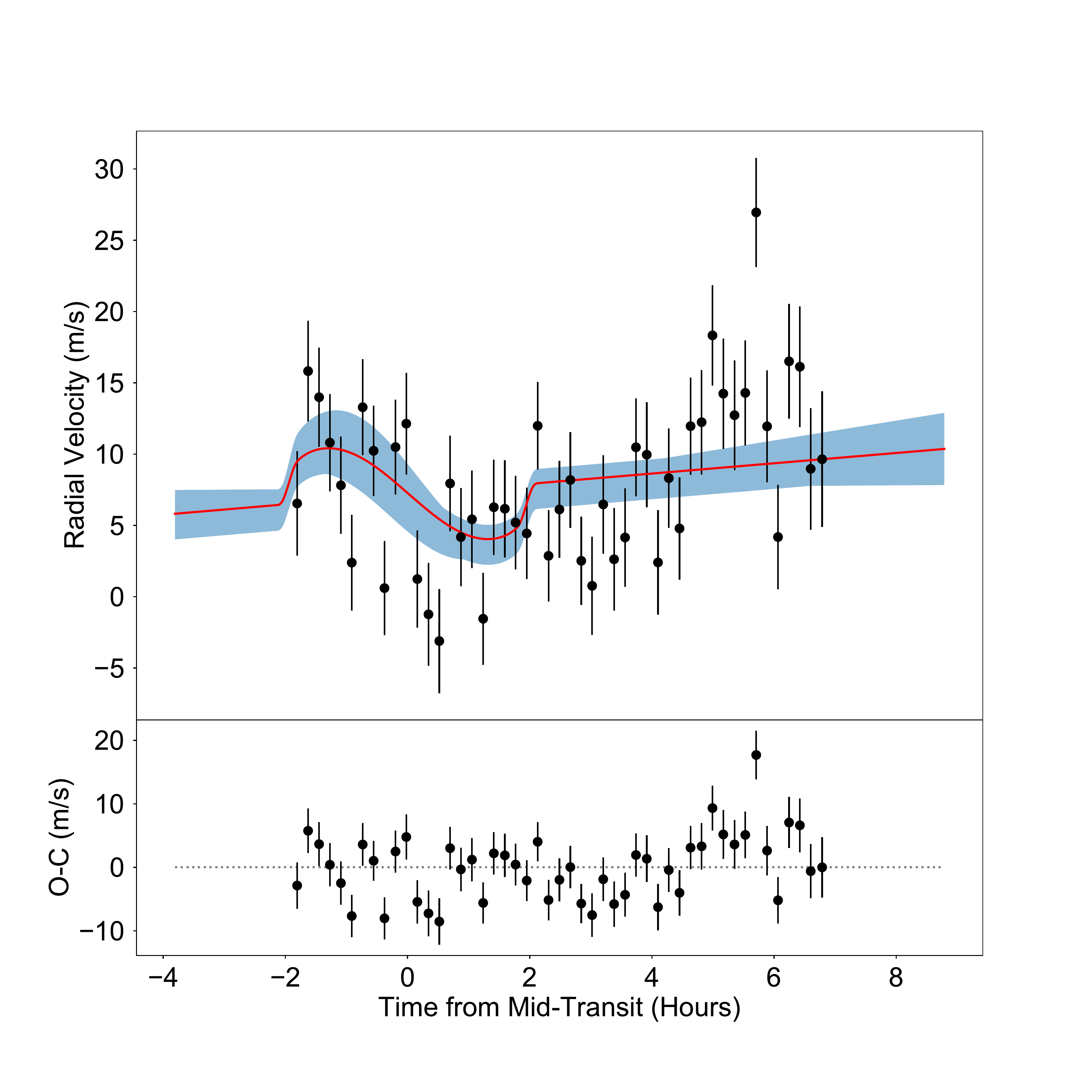}
\caption{The measured radial velocities during the transit of TOI-1726 c. The red line is the best-fit model; the blue shaded region represent the 68\% confidence region from the posterior distribution. The data suggest a stellar obliquity of  $-1^{+35}_{-32}~^{\circ}$ that favors a prograde, and likely aligned orbit for TOI-1726 c . Visually, there are also hints of a red noise component towards the end of observation. We investigated the source of this red noise component with line profile analysis and its effect on obliquity measurement with a Prayer's Beads analysis in Section 4.}
\label{fig:fit}
\end{center}
\end{figure*}

\begin{figure*}
\begin{center}
\includegraphics[width = 0.2873\columnwidth]{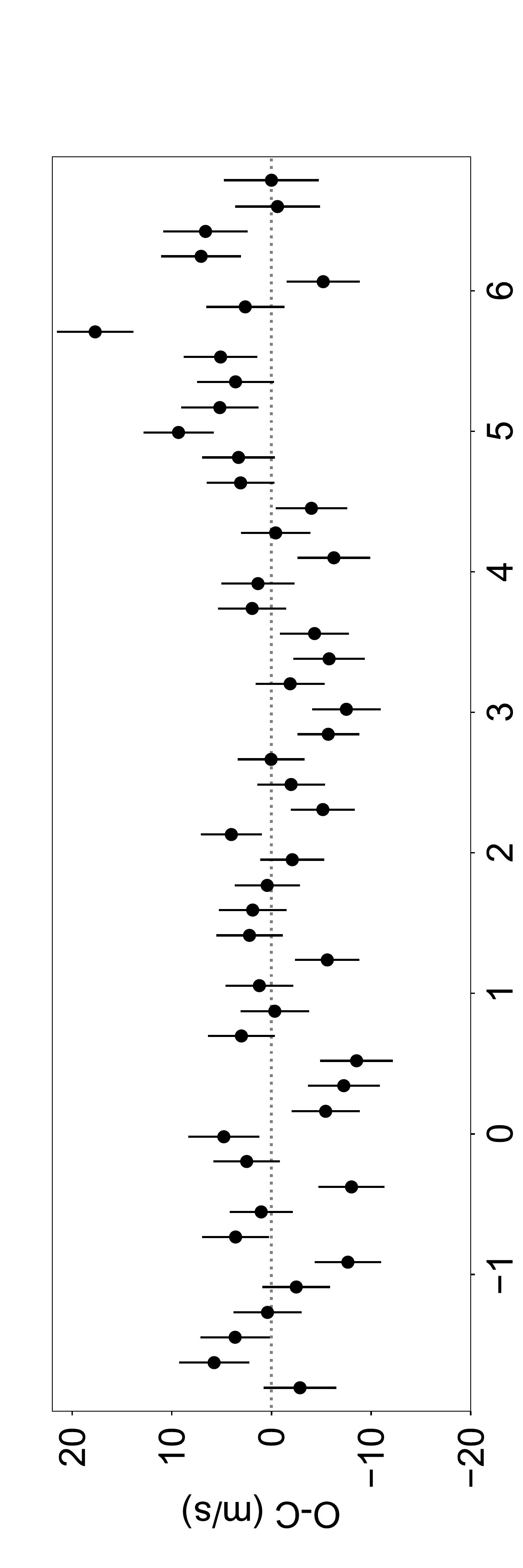}
\includegraphics[width = 1.122\columnwidth]{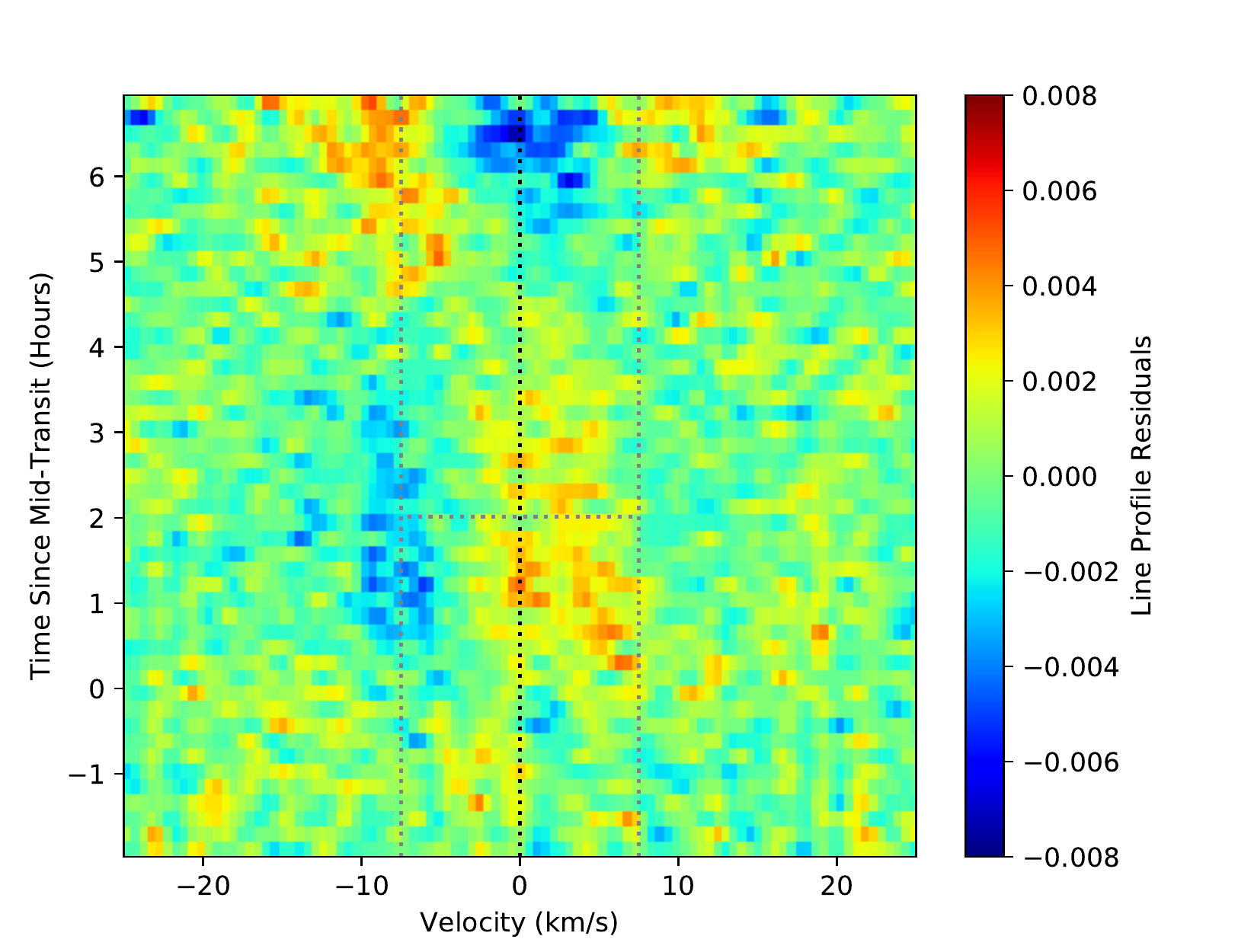}
\includegraphics[width = 1.122\columnwidth]{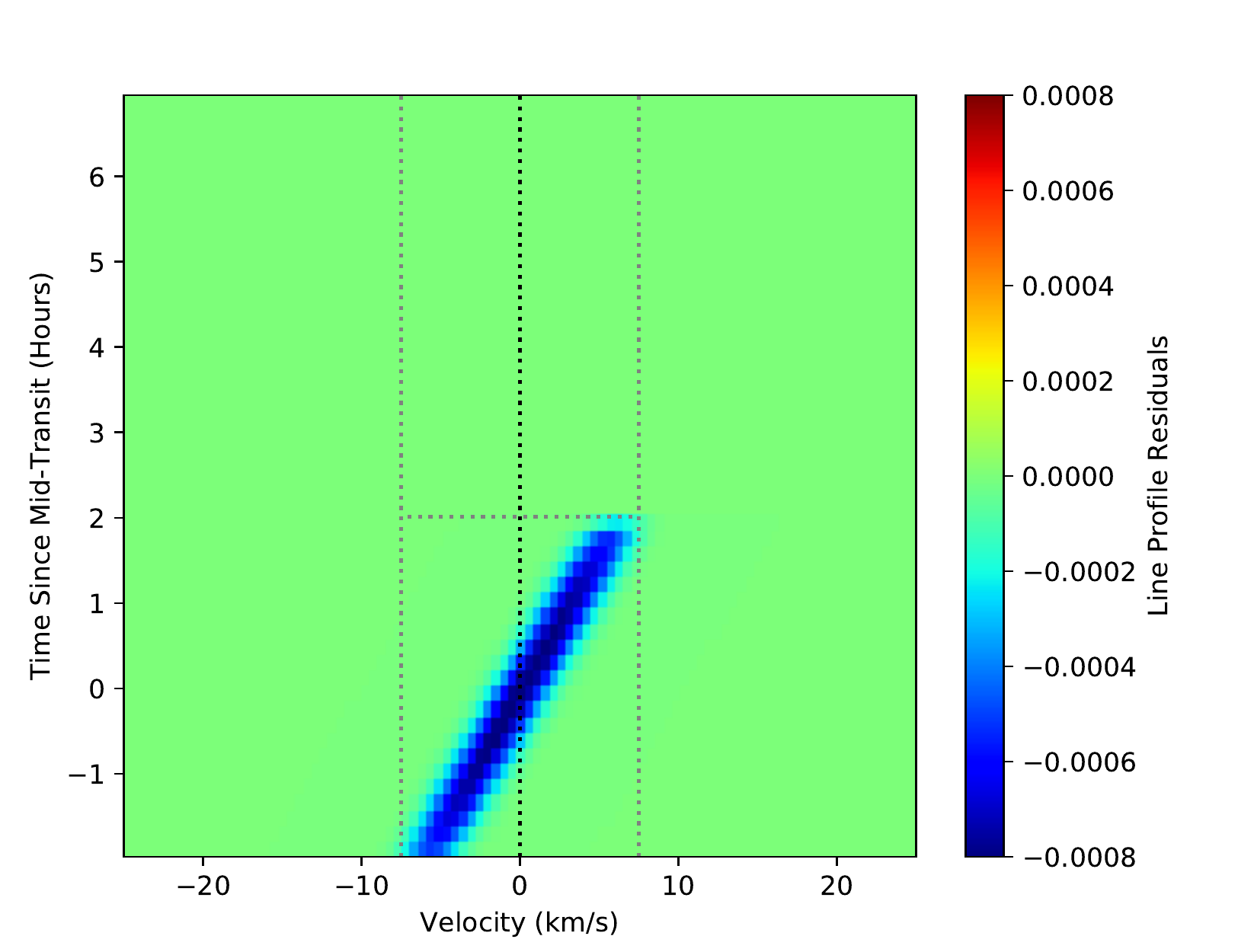}
\caption{{\bf Top Left}: The residuals of the RM time series same as Figure 1. {\bf Top Right}: The measured line profile residuals as a function of time and velocity. The vertical gray lines indicate the vsin$i$ of the host star. The horizontal gray line indicate the end of the transit $t_{\rm IV}$. Some localized patterns can be seen which are likely due to a combination of stellar activity and instrumental drifts. {\bf Bottom}: The simulated planetary shadow of TOI-1726c on a well-aligned orbit. The signal is about one order of magnitude lower than the uncertainties seen in the measurements (note the different color coding in these two panels); and remains undetected with the current measurement.}
\label{fig:dt}
\end{center}
\end{figure*}

\begin{figure*}
\begin{center}
\includegraphics[width = 2.3\columnwidth]{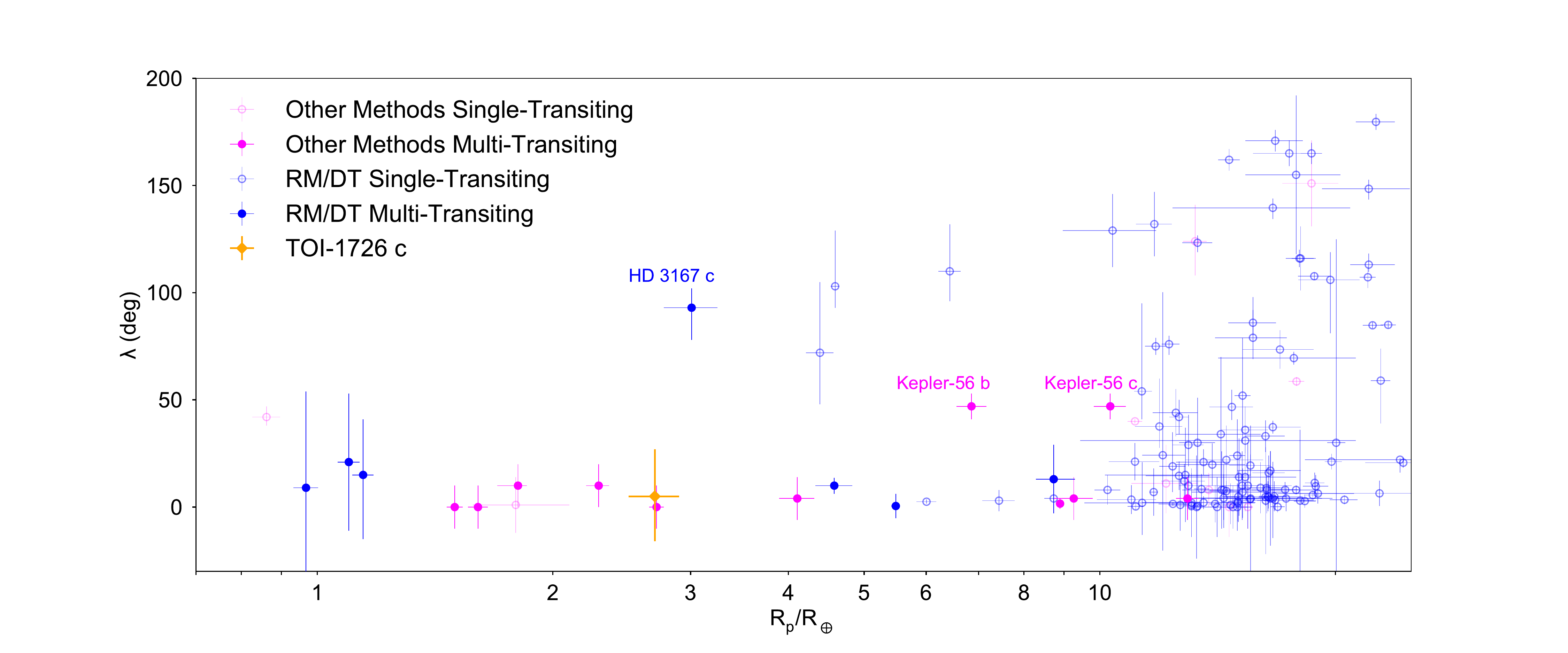}
\includegraphics[width = 2.3\columnwidth]{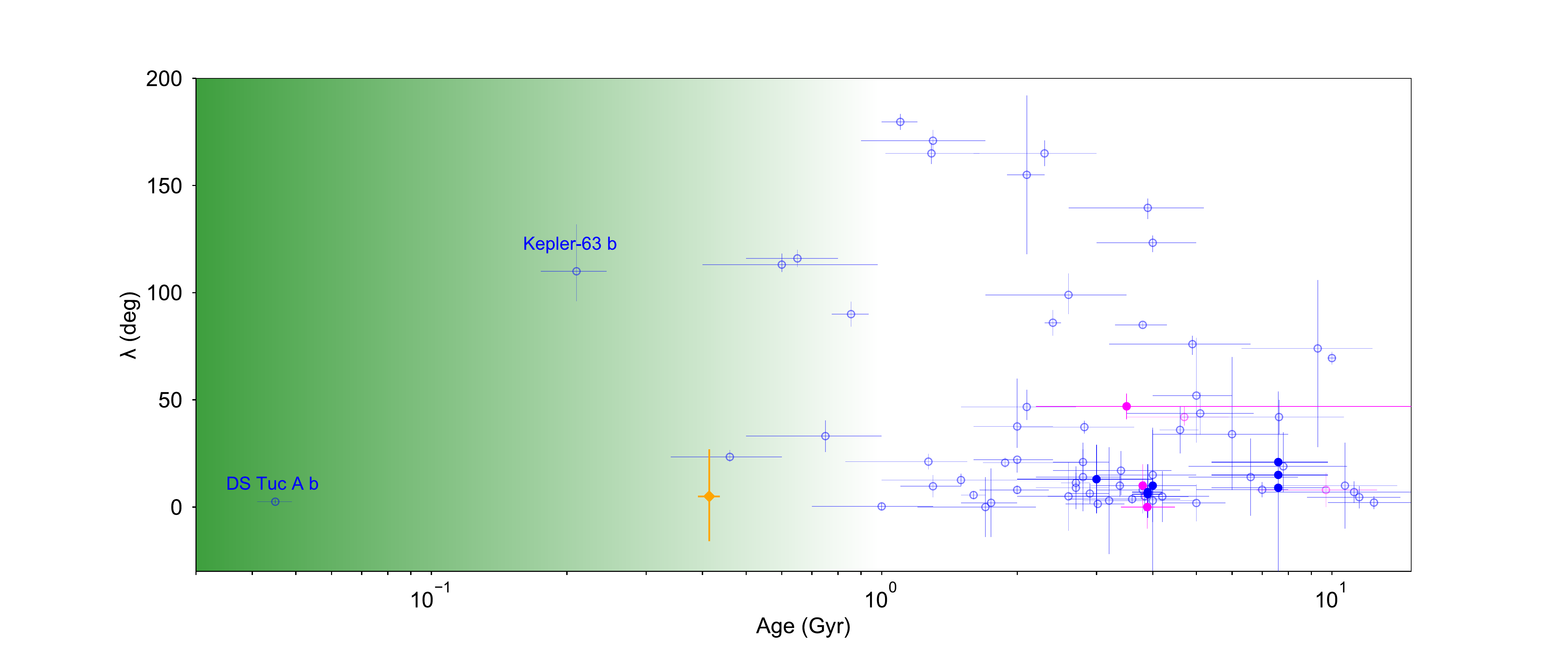}
\caption{The projected stellar obliquity $\lambda$ plotted against the planetary radius (Upper) and stellar age (Lower). The majority of stellar obliquity measurements are performed for single-transiting planets which are believed to have a dynamically hot history. We highlighted measurements of relatively unexplored multi-transiting systems with filled symbols. TOI-1726 is a unique opportunity for obliquity measurement for multi-transiting sub-Neptune planetary systems with a well-determined young age. The green shading in the lower panel qualitatively captures the magnitude of the high-energy radiation from the host star that is responsible for driving photoevaporation. These high-energy radiation dwindles with the first few hundred Myr: a timescale future observations of TOI-1726 are poised to probe.}
\label{fig:lam}
\end{center}
\end{figure*}

\section{Spectroscopic Measurement}
We obtained 49 spectra of TOI-1726 on the night of UTC 2020 Feb 26, spanning a transit of TOI-1726 c.
We used the Automated Planet Finder \citep[APF,][]{Vogt} at the Lick Observatory. The spectra were obtained with an iodine cell whose dense forest of molecular lines provide both the wavelength solution and a means of determining the line spread function. The spectral resolution was $\sim$100,000. We obtained consecutive 10-min exposures that enabled a median SNR of 145 per reduced pixel near 5500 Å.
The iodine method for determining precise radial velocities requires a template spectrum of the star
with a high signal-noise-ratio.  A template spectrum should have been obtained with APF. However, due to weather conditions and scheduling constraints, we had to secure a high SNR template of TOI-1726 on the High Resolution Echelle Spectrometer on the 10m Keck I telescope on the night of UT 2020 Mar 10. APF and HIRES have similar instrumental designs \citep{HIRES,APF}. Moreover, we explicitly deconvolved the different instrumental profiles from the template spectra in our Doppler pipeline \citep{Howard}, therefore the high SNR, iodine-free HIRES spectrum should serve adequately as the template spectrum for reducing the APF dataset. More details of our forward-modeling Doppler pipeline are described in \citet{Howard}.  The radial velocities and uncertainties are plotted in Fig. \ref{fig:fit} and reported in Table \ref{tab:rv}.

\section{Stellar Parameters}
\label{sec:stellar}
We constrained the spectroscopic parameters ($T_{\rm eff}$, log $g$ and [Fe/H]) of TOI-1726 using the iodine-free spectra from Keck/HIRES and the SpecMatch pipeline\footnote{\url{https://github.com/petigura/specmatch-syn}} \citep{CKS1}. In short, SpecMatch models observed optical spectra with interpolated model spectra from the precomputed grid \citep{Coelho} of discrete $T_{\rm eff}$, [Fe/H], log$~g$ and $v$sin$i$ values. Line broadening effects from both rotation and macroturbulence are included by convolving the model spectra with the kernel described by \citet{Hirano}. Instrumental broadening is modeled as a Gaussian function with a FWHM of 3.8 pixels, a value that provides a good match to the widths of telluric lines. We calculate the weighted average of spectroscopic parameters of five $\sim400$ Å spectral segments. The final output spectroscopic parameters are corrected for known systematic effects from previous comparison with standard stars. Particularly, SpecMatch systematically yields higher ($\sim$ 0.1 dex) surface gravity log $g$ for earlier-type stars when compared with asteroseismic results of standard stars \citep{Huber2013}. This effect is empirically corrected for with a scaling relation log $g$($T_{\rm eff}$,[Fe/H]). See \citet{Petigura_thesis} for detail.

To derive the stellar parameters, we further make use of Gaia parallax information \citep{Gaia}. We followed the procedure described in detail by \citet{CKS7}. To summarize, we link the stellar effective temperature, the parallax measurement from Gaia and the $K$-band magnitude (which is less affected by extinction) together with Stefan–Boltzmann Law for an independent constraint on the radius of the star. In practice, we put in the priors on spectroscopic parameters and the parallaxes into the {\it Isoclassify} package of \cite{Huber} which then compares these parameters with the MESA Isochrones \& Stellar Tracks \citep[MIST,][]{MIST} to determine the posterior distribution of various stellar parameters. The results are summarized in Table \ref{tab:para}.

\section{Joint Light Curve and RM Analysis}
TOI-1726 was observed by {\it TESS} \citep{Ricker} in Sector 20 from UT 2019 Dec 24 to 2020 Jan 20. We downloaded the reduced light curve from the Mikulski Archive for Space
Telescopes website\footnote{\url{https://archive.stsci.edu}}. We only kept data points with a Quality Flag of 0, i.e., those with no known problems.

 We started from the transit ephemerides reported by the {\it TESS} team. We first removed the data spanning the transits of both planet b and c from the light curve. This enabled us to measure the stellar rotation period of TOI-1726 by applying the Lomb-Scargle periodogram. We detected a strong rotational modulation at a period of $6.36^{+0.75}_{-0.25}$ days where the uncertainties are derived from the full width half maximum of the peak. As a consistency check, we calculated $v=2\pi R_\star/P_{\rm rot}$, the rotation period of $6.36^{+0.75}_{-0.25}$ days and the stellar radius of $0.92\pm-0.10R_\odot$ together give a rotational velocity $v$ of $7.3^{+0.7}_{-1.0}$ km/s which is consistent with the $v$sin$i$ of $6.56 \pm 1.0$ km/s determined from the spectroscopic analysis alone. Using the procedure described in \citet{Masuda_vsini}, the orbital inclination of the host star is >45$^\circ$ at 95\% confidence level. This agreement of  $v$sin$i$ and $v$ is supporting evidence for a prograde and perhaps well-aligned orbit of TOI-1726 c, in addition to the analysis of the RM effect described later in the paper.
 
 We then analyzed the in-transit light curve simultaneously with the Rossiter-McLaughlin effect. We isolated data taken within one transit duration of the transit midpoint.
We used the {\tt Batman} package \citep{Kreidberg2015} to model the transit light curves.
We adopted a quadratic limb-darkening law, imposing Gaussian priors on the coefficients
with medians taken from precomputed limb darkening coefficients from {\tt EXOFAST}\footnote{\url{astroutils.astronomy.ohio-state.edu/exofast/limbdark.shtml}.} \citep{Eastman2013} and with widths of 0.3. We put a prior on the mean stellar density based on the analysis in Section \ref{sec:stellar}. We sampled $P_{\rm orb}$, $R_{\rm p}/R_\star$ and $a/R_\star$ uniformly in logarithmic space. We put a uniform prior on the impact parameter $b$ [-1,1] and on the midtransit time ($T_{\rm c}$). We assumed that both planets are on circular orbits. The current RV dataset (Hirsch et al. in prep) only provide weak constraints on the orbital eccentricities and are consistent with being circular for both planets.

To model the RM effect, we followed the prescription of \citet{Hirano}. The additional parameters are the sky-projected obliquity $\lambda$, the projected rotational velocity  $v$sin$i$ the radial velocity offset $\gamma$ and the local gradient of the offset $\dot{\gamma}$. We also included a jitter parameter to account for any additional astrophysical or instrumental noise. The likelihood function of the RM model was combined with the likelihood function of the transit model. 

We sampled the posterior distribution using the Markov Chain Monte Carlo technique
implemented in the {\textit emcee} code \citep{emcee}. We used 128 walkers and ran until the Gelman-Rubin convergence statistics dropped below 1.03. We first included a prior on the rotational modulation $v$sin$i$ of $6.56 \pm 1.0$ km/s from spectroscopic analysis in Section \ref{sec:stellar}. The sky-projected obliquity has a posterior distribution of $-6^{+28}_{-25}~^{\circ}$ i.e. favor a prograde and possibly aligned orbit for planet c. The posterior distribution also favors a slightly higher $v$sin$i$ of $7.0 \pm 1.0$ km/s. When we removed the prior on $v$sin$i$ altogether, the data are consistent with a broader range of $v$sin$i$ of $9.9^{+4.3}_{-3.4}$ km/s; while the posterior distribution of stellar obliquity also widened $\lambda$ $-5^{+33}_{-26}~^{\circ}$.
Table \ref{tab:para} reports the summary of the posterior distribution for the key parameters.

\section{Doppler Tomography and Red Noise Mitigation}
 We tried to look for the Doppler shadow of planet c in the subtle variation of the line profiles using the non-Iodine part of the spectra ($4000-5000$ Å). Our analysis is similar to that of \citet{Albrecht}. In short, we cleaned the spectrum from outliers with 5-sigma clipping. We removed the continuum and blaze function with a polynomial fit to the 95$\%$ percentile flux level in each Echelle order. We cross-correlated the individual spectrum with the bestfit SpecMatch spectrum before rotational/instrumental broadening is applied. We then subtracted the globally averaged line profile from the individual line profiles to extract the subtle variations that may be caused by the shadow of the transiting planet (Figure 2).  However, given the small transit depth of the planet ($\sim$0.08\%), we could not convincingly detect the shadow of the planet in the line profile residuals. Instead, the line profile residuals are dominated by patterns that are almost one order of magnitude larger in amplitude; roughly constant in velocity and extend well beyond the transit duration. We suspect that these pattern most likely produced by the change of the point spread function due to instrumental effects or the emergence of stellar activity on TOI-1726. However, we do not have a physically-motivated model to eliminate these effects.

Visual inspection of the residuals of the RM time series hint at the presence of a correlated noise component more noticeably starting at 4 hours after the mid-transit of planet c (Figure 1). This coincided the onset of correlated noises in the radial velocity residuals (Figure 2) as well as an increase of the S index (Table 1). To assess how the presence of a correlated noise component might have affected the constraint on stellar obliquity, we performed a Prayer's Beads analysis. This is perhaps more worrisome as some of the line profile residual patterns happened during the transit of planet c (Figure 2). Our analysis is as follows. We first found the maximum likelihood model with the Levenberg-Marquardt method as implemented in {\sc Python} package {\sc lmfit}. We recorded the corresponding residuals and cyclically permuted the residuals before adding them back to the best-fit model. This generated a series of mock datasets that contains the same correlated noise component as the original dataset. We found the maximum likelihood model for each mock dataset. Focusing on the stellar obliquity, the resultant distribution of obliquity is $\lambda =-1^{+35}_{-32}~^{\circ}$. This is a broader distribution compared to that from the white-noise-only model in Section 4; but qualitatively these two models both favor a prograde, possibly aligned orbit for TOI-1726c.

\section{Discussion}
\subsection{Obliquity of Multi-Transiting Systems}
It has been noted in several previous works that the underlying orbital architectures of {\it Kepler} single-transiting (here we refer to the observed multiplicity, to be distinguished from planets that only transited host stars once during the time span of observation) and multi-transiting systems may be different. Specifically, single-transiting systems seem to have a broader distribution of orbital eccentricities whereas multi-transiting systems mostly favor circular orbits \citep{VanEylen_multi2015,Xie,Mills2019}. In addition, \citet{Fang2012} and \citet{Zhu2018} suggested that the mutual inclination dispersion is larger when the observed multiplicity of a planetary system is smaller. A plausible explanation of this architectural difference is the dynamical interaction between the sub-Neptune planets or that with a more distant giant planets. \citet{ZhuSE-CJ} and \citet{Bryan} independently arrived at the conclusion that {\it Kepler}-like sub-Neptune planets are much more likely to have a cold Jupiter companion (>1AU) than randomly chosen stars \citep{Cumming,Clanton}. \citet{Masuda} further showed that when the inner planetary system only has one transiting planet, its cold Jupiter is likely inclined by tens of degrees relative to the inner planetary system. The interpretation is that the dynamical interaction of an inclined cold Jupiter can stir up the initially co-planar planetary systems while exciting larger mutual inclinations and eccentricities. The single-transiting systems represent the dynamically hot sub-sample while the multi-transiting systems are dynamically colder.

It will be interesting to see if the same architectural difference carries over to the stellar obliquity distribution. So far, there are about 150 obliquity measurements in the literature.  Traditional RM effect is more easily detected for planets with larger radii and more frequent transits. As a result, the vast majority of existing measurements were performed for hot Jupiters or hot Neptunes. Intriguingly, it is often the case that these hot Jupiters and hot Neptunes are single-transiting planets with spin-orbit misalignments both of which hint at a dynamically hot past \citep{Dong}. On the other hand, multi transiting systems tend to display low obliquities \citep{Albrecht}. Unfortunately there are only $\sim 11$ obliquity measurements obtained for multi-transiting systems to date (see Fig.~\ref{fig:lam}). We note the most complete census of spin-orbit angle of multi-transiting systems was done by \citet{Winn}. They compared the projected rotational velocity $v$sin$i$ and the rotational velocity $v=2\pi R_\star/P_{\rm rot}$. If a system is grossly misaligned, $v$sin$i$ would be much smaller than $v$. \citet{Winn} found that the majority of {\it Kepler}-like systems (systems with several sub-Neptune planets within 1AU) are well-aligned with their host star. The six high-obliquity suspects \citet{Winn} identified were dominated by hot Jupiters. This result revealed a picture that planets in multi-planet systems are generally well-aligned as one would expect from a cold dynamical history. Coming back to the multi-transiting systems that have their stellar obliquities explicitly measured, most of these measurements were often obtained with alternative methods, rather than the RM effect, such as asteroseismology \citep[e.g.][]{Huber+2013} or spot-crossing anomalies \citep[e.g.][]{Sanchis}. The results mostly yield well-aligned orbits. We note that the only exceptions are the polar orbit of HD~3167\,c \citep{HD3167} and 50$^{\circ}$ inclined orbit of Kepler-56 b and c \citep{Huber+2013}.  What kind of formation channel gave rise to misaligned multi-planet systems have been a topic of interests for the theorists \citep[e.g.][]{Li,Spalding2015}. It will be interesting to see if these two systems are indeed rare occurrences. Our result on TOI-1726 c is one crucial step towards enlarging that sample of multi-transiting planetary system.  Although the obliquity constraints on planet b and planet c individually are weak: $1^{+41}_{-43}~^{\circ}$ \citep{Mann2020} versus $-1^{+35}_{-32}~^{\circ}$, the fact that both planets transit and posterior distribution of obliquity both center at 0 seems to favor a coplanar, likely aligned, dynamically quiet architecture for TOI-1726.

\subsection{Obliquity in Time}
As we mentioned briefly in the introduction, many different theories have been offered to explain the observed diversity of stellar obliquities \citep[e.g.][]{Fabrycky,WuLithwick,Lai,Batygin,Yee}. Since these theories operate on very different timescales, a potential way to test some of them is to obtain obliquity measurements for a sample of planets with well-determined ages. For example, if young planetary systems rarely display spin-orbit misalignment, it is reasonable to say that the orbit-tilting mechanisms that only operate during the disk-hosting stage \citep[e.g.][]{Lai,Batygin} cannot be the dominant channel to generate spin-orbit misalignment. The cluster membership of TOI-1726 \citep{Mann2020} provides a firm and precise age estimate for the host star. In Fig.~\ref{fig:lam}, we plotted all obliquity measurements for systems with better than 20\% age estimates. TOI-1726 c is the third youngest planet with obliquity measurement. Moreover TOI-1726 c is a sub-Neptune which is the predominant product of planet formation in the Galaxy \citep{Petigura}, whereas a group of planets for which obliquity measurements have been lacking (Fig. \ref{fig:lam}).

\subsection{A great system for studying atmospheric losses}

The bimodal radius distribution and the presence of the so-called "Hot Neptune Desert" both suggest that atmospheric loss from sub-Neptune planets is a common if not ubiquitous phenomena \citep{Fulton}. TOI-1726 is a great system for a study of atmospheric loss. The star is 400 Myr old which is comparable to the timescale where  high-energy radiation from the host star begins to diminish \citep{Ribas} and the photoevaporation starts to come to a conclusion (see Fig. \ref{fig:lam}). Moreover, the system contains two sub-Neptune planets whose low surface gravity make them the planets most amenable to photoevaporation \citep{Wang}. The two planets are suited to comparative study since they orbit around the same host star. In other words, the planets are bathed in the same high-energy radiation environment except for a difference in orbital distance. Any difference in the outcome of atmospheric loss has to come from the different planetary parameters e.g. orbital period and planetary mass etc. The prograde and coplanar orbits of both planet b \citep{Mann2020} and planet c together disfavor a violent event such as high-eccentricity migration or giant impact collision that would have disrupted the planets' coplanarity and complicated the evolution of the atmospheres. We also note that there is no compelling evidence for a cold Jupiter that may generate dynamical instability of the inner planetary system ($\sim$8000-day baseline, Hirsch et al. in prep).

\acknowledgements
We thank the time assignment committee of the University of California for observing time on the Automated Planet Finder for the TESS-Keck Survey. We thank NASA for funding associated with our Key Strategic Mission Support project for TKS.  We thank Ken and Gloria Levy, who supported the construction of the Levy Spectrometer on the Automated Planet Finder. We thank the University of California and Google for supporting Lick Observatory and the UCO staff for their dedicated work scheduling and operating the telescopes of Lick Observatory. This paper is based on data collected by the TESS mission. Funding for the TESS mission is provided by the NASA Explorer Program.  T.\,D. acknowledges support from MIT's Kavli Institute as a Kavli postdoctoral fellow. M.\,N\,G acknowledges support from MIT's Kavli Institute as a Torres postdoctoral fellow. P.\,D. acknowledges support from a National Science Foundation Astronomy and Astrophysics Postdoctoral Fellowship under award AST-1903811. JMAM gratefully acknowledges support from the National Science Foundation Graduate Research Fellowship under Grant No. DGE-1842400. JMAM also thanks the LSSTC Data Science Fellowship Program, which is funded by LSSTC, NSF Cybertraining Grant No. 1829740, the Brinson Foundation, and the Moore Foundation; his participation in the program has benefited this work.MHK anknowledges Allan R. Schmitt for making his lightcurve examining software LcTools freely available. DC was supported by a grant from the John Templeton Foundation. The opinions expressed in this publication are those of the authors and do not necessarily reflect the views of the John Templeton Foundation. SA acknowledge the support from the Danish Council for Independent Research through the DFF Sapere Aude Starting Grant No. 4181-00487B, and the Stellar Astrophysics Centre which funding is provided by The Danish National Research Foundation (Grant agreement no.: DNRF106)

\facilities{Automated Planet Finder (Levy), \textit{TESS}}

\software{Batman \citep{Kreidberg2015},
          Emcee \citep{emcee},
          EXOFAST \citep{Eastman2013},
          Isoclassify \citep{Huber17}, lmfit\citep{LM}
          SpecMatch \citep{Petigura15, Yee17},
          }

\bibliography{main}

\begin{deluxetable}{ccccc}
\tabletypesize{\scriptsize}
\tablecaption{Lick/APF Radial Velocities \label{tab:rv}}
\tablehead{
\colhead{Time (BJD)} & \colhead{RV (m/s)} & \colhead{RV Unc. (m/s)} & \colhead{S index}& \colhead{S Unc.} }
\startdata
2458905.618603&6.54&3.66&0.374&0.002\\
2458905.626056&15.82&3.52&0.381&0.002\\
2458905.633579&13.99&3.47&0.373&0.002\\
2458905.640951&10.81&3.41&0.372&0.002\\
2458905.648485&7.82&3.41&0.375&0.002\\
2458905.655846&2.39&3.35&0.377&0.002\\
2458905.663311&13.29&3.37&0.382&0.002\\
2458905.670752&10.23&3.17&0.385&0.002\\
2458905.678159&0.61&3.31&0.375&0.002\\
2458905.685705&10.49&3.33&0.384&0.002\\
2458905.693031&12.14&3.57&0.386&0.002\\
2458905.700588&1.24&3.41&0.389&0.002\\
2458905.708180&-1.23&3.61&0.381&0.002\\
2458905.715564&-3.11&3.65&0.382&0.002\\
2458905.722925&7.94&3.36&0.380&0.002\\
2458905.730262&4.18&3.45&0.383&0.002\\
2458905.737843&5.43&3.42&0.381&0.002\\
2458905.745470&-1.55&3.22&0.379&0.002\\
2458905.752749&6.28&3.35&0.385&0.002\\
2458905.760260&6.17&3.41&0.381&0.002\\
2458905.767575&5.20&3.28&0.381&0.002\\
2458905.775190&4.44&3.20&0.391&0.002\\
2458905.782655&11.99&3.08&0.380&0.002\\
2458905.790061&2.87&3.21&0.384&0.002\\
2458905.797492&6.12&3.40&0.384&0.002\\
2458905.804945&8.19&3.35&0.381&0.002\\
2458905.812386&2.52&3.11&0.379&0.002\\
2458905.819793&0.77&3.44&0.367&0.002\\
2458905.827362&6.47&3.46&0.391&0.002\\
2458905.834758&2.63&3.59&0.382&0.002\\
2458905.842257&4.15&3.46&0.378&0.002\\
2458905.849699&10.48&3.43&0.379&0.002\\
2458905.857059&9.96&3.69&0.381&0.002\\
2458905.864721&2.41&3.66&0.375&0.002\\
2458905.872093&8.32&3.48&0.380&0.002\\
2458905.879396&4.79&3.59&0.374&0.002\\
2458905.886930&11.96&3.41&0.380&0.002\\
2458905.894476&12.24&3.67&0.376&0.002\\
2458905.901883&18.33&3.53&0.384&0.002\\
2458905.909255&14.24&3.88&0.385&0.002\\
2458905.916893&12.73&3.85&0.382&0.002\\
2458905.924254&14.29&3.69&0.383&0.002\\
2458905.931707&26.95&3.83&0.387&0.002\\
2458905.939126&11.95&3.92&0.386&0.002\\
2458905.946590&4.19&3.66&0.394&0.002\\
2458905.954113&16.51&4.02&0.383&0.002\\
2458905.961497&16.13&4.24&0.388&0.002\\
2458905.968869&8.97&4.26&0.383&0.002\\
2458905.976681&9.64&4.76&0.391&0.002\\
\enddata
\end{deluxetable}

\begin{deluxetable}{lll}
\tabletypesize{\scriptsize}
\tablecaption{Stellar and Transit Parameters of planet c \label{tab:para}}
\tablehead{
\colhead{Parameter}  & \colhead{Symbol} &  \colhead{Posterior Distribution} }
\startdata
Sky-projected Obliquity (deg)& $\lambda$  & $-1^{+35}_{-32}$ \\
Projected Stellar Rotation (km/s)& $v$sin$i$  & $7.0\pm1.0$ \\
Radial Velocity Offset (m/s)& $\gamma$  & $4.69^{+1.95}_{-1.98}$ \\
Radial Velocity Trend (m/s/day)& $\dot{\gamma}$  & $19.4^{+8.5}_{-8.8}$ \\
Planet/Star Radius Ratio & $R_p/R_\star$  & $0.02660^{+0.00082}_{-0.00074}$  \\
Planetary Radius ($R_\oplus$)  & $R_{\rm p}$ &$2.71\pm0.14$ \\
Time of Conjunction (BJD-2457000) & $t_0$  & $1844.0577\pm0.0011$  \\
Impact Parameter & $b$  & $0.50\pm0.07$ \\
Scaled Semi-major Axis & $a/R_\star$  & $38.0^{+1.7}_{-4.6}$  \\
Orbital Period (days) & $P_{\rm orb}$   & $20.5456^{+0.0016}_{-0.0019}$  \\
Jitter (m/s)  & $\sigma$ &$3.66^{+0.80}_{-0.75}$ \\
Effective Temperature ($T_{\rm eff}$)  & $K$ &$5710\pm100$ \\
Surface Gravity (dex)  & log$~g$ &$4.6 \pm 0.1$ \\
Metallicity (dex)  & [Fe/H] &$0.05 \pm 0.05$ \\
Projected Stellar Rotation from Spectroscopy (km/s)& $v$sin$i$  & $6.56\pm1.0$ \\
Stellar Mass ($M_\odot$)  & $M_\star$ &$0.994\pm0.036$ \\
Stellar Radius ($R_\odot$)  & $R_\star$ &$0.934\pm0.019$ \\
Stellar Density (g/cm$^3$)  & $\rho_\star$ &$1.72\pm0.17$ \\
Rotation Period (days)  & $P_{\rm rot}$ & $6.36^{+0.75}_{-0.25}$ \\
\enddata
\end{deluxetable}

\end{document}